\documentclass{article}

\usepackage{PRIMEarxiv}
\usepackage{multicol}
\usepackage{xcolor}
\usepackage[utf8]{inputenc} 
\usepackage[T1]{fontenc}    
\usepackage{hyperref}       
\usepackage{url}            
\usepackage{booktabs}       
\usepackage{amsfonts}       
\usepackage{nicefrac}       
\usepackage{microtype}      
\usepackage{lipsum}
\usepackage{fancyhdr}       
\usepackage{graphicx}       
\usepackage{float}
\usepackage{subcaption}
\usepackage{caption}
\usepackage{adjustbox}
\usepackage{makecell}
\usepackage{tikz}
\usepackage{pgfplots}
\usepackage{amsmath}
\usepackage{svg}
\usepackage{multirow}
\usepackage{enumitem}

\title{Concordance in basal cell carcinoma diagnosis. Building a proper ground truth to train Artificial Intelligence tools}

\author{ {Francisca Silva-Clavería, MD} \\
         Servicio de Dermatología \\
         Hospital Universitario Virgen Macarena \\
         Seville, Spain \\
         \And
         {Carmen Serrano, PhD} \\
         Dpt. Teoría de la Señal y Comunicaciones \\
         Escuela Técnica Superior de Ingeniería \\
         Universidad de Sevilla, Seville, Spain \\    	
         \AND
         {Iván Matas, MD} \\
         Dpt. Teoría de la Señal y Comunicaciones \\
         Escuela Técnica Superior de Ingeniería \\
         Universidad de Sevilla, Seville, Spain \\
         \And
         {Amalia Serrano, MD} \\
         Servicio de Dermatología \\
         Hospital Universitario Virgen Macarena\\
         Seville, Spain \\
         \And
         {Tomás Toledo-Pastrana, PhD} \\
         Hospitales Quironsalud Infanta Luisa \\
         y Sagrado Corazón \\
         Seville, Spain \\
         \And
         {Begoña Acha, PhD} \\
         Dpt. Teoría de la Señal y Comunicaciones \\
         Escuela Técnica Superior de Ingeniería \\
         Universidad de Sevilla, Seville, Spain \\
        }

\begin{document}
    
    \maketitle

    \begin{abstract}
        \textbf{Background}: The existence of different clinical criteria for basal cell carcinoma (BCC) cannot be objectively validated. An adequate ground truth is needed to train an artificial intelligence (AI) tool that explains the diagnosis of BCC by providing its dermoscopic features.
        
        \textbf{Objectives}: Determine the consensus among dermatologists on the dermoscopic criteria for 204 BCC. To analyze the performance of an AI tool when the ground-truth is inferred.
        
        \textbf{Methods}: A single-center diagnostic and prospective study was conducted to analyze agreement in dermoscopic criteria by four dermatologists and then derive a reference standard. 1434 dermoscopic images have been used, that were taken by a primary health physician, sent via teledermatology, and diagnosed by a dermatologist. They were randomly selected from the teledermatology platform (2019-2021). 204 of them were tested with an AI tool; the remainder trained it. The performance of the AI tool trained using the ground truth of one dermatologist versus the ground-truth statistically inferred from the consensus of four dermatologists was analyzed using McNemar's test and Hamming distance.
        
        \textbf{Results}: Dermatologists achieve perfect agreement in the diagnosis of BCC (Fleiss-Kappa = 0.990), and a high correlation with biopsy (PPV = 0.9670). However, there is low agreement on detecting some dermoscopic criteria. Statistical differences were found in the performance of the AI tool trained using the ground truth of one dermatologist versus the ground-truth statistically inferred from the consensus of four dermatologists. 
        
        \textbf{Conclusions}: Care should be taken when training an AI tool to determine the BCC patterns present in a lesion. The ground truth should be established from multiple dermatologists.
        
        \end{abstract}


        \section{Introduction}\label{sec:Introduction}
        Skin cancer is the most common cancer worldwide \cite{skincancerfoundation2024}. There are two main types of skin cancer: melanoma and non-melanoma. The most frequent non-melanoma tumors are basal cell carcinoma (BCC) with an incidence over 70\% \cite{aad2024}, with the best validated clinical criteria for diagnosis and the greatest variability in the presence of these clinical criteria \cite{peris2019diagnosis}.
        Numerous papers in the scientific literature apply artificial intelligence techniques to aid in the diagnosis of skin diseases. Especially, that number has increased considerably in recent years due to the existence of public databases \cite{Codella2017, combalia2019bcn20000, HAM10000DataBase, isicarchive2024}. These databases are accessible and comprehensive, but the clinical criteria that have been used in the diagnosis of the lesions are not available. For a tool to be useful from a medical point of view, we consider it crucial to provide not only the classification of the lesion, but also an explained diagnosis with the detected clinical features that motivates this classification to develop a clinically useful tool.
        In this regard, Serrano et al \cite{serrano2022} developed a clinically inspired skin lesion classification tool through the detection of dermoscopic criteria for BCC.
        However, a suitable Ground Truth (GT) is needed to train an Artificial Intelligence (AI) tool focused on inferring the dermoscopic structures present in a lesion. 
        Rodríguez-Lomba et al attempt to illustrate the subjectivity present in the detection of different dermoscopic features \cite{rodriguez2022dermal}. To this aim, they calculate the concordance between five dermatologists in detecting twenty-two features and obtain a Kappa-Fleiss agreement coefficient ranging from 0.04 and 0.46, indicating slight or fair agreement \cite{rodriguez2022dermal, landis1977measurement}. 
        This lack of agreement has prompted researchers to avoid determining the GT or reference standard\cite{lallas2014dermoscopy} from a single observer. For example, Longo et al. estimate the BCC dermoscopic criteria present in a lesion from two observers, and when they disagree, a third observer breaks the tie12. In recent years, attention has been paid to the determination of a GT from different raters, especially in the context of crowdsourcing \cite{zhang2016learningCrowdsourced}.
        In this paper, four dermatologists determine the BCC dermoscopic criteria present in 204 lesions. On one hand, consensus among dermatologists is analyzed. On the other hand, the GT is inferred from the diagnosis of the four different raters. Furthermore, how the GT affects the performance of a supervised AI tool is also analyzed.

        \section{Material and methods}
        \subsection{Database}
\begin{figure}[ht]
            \centering
            \includegraphics[width = \textwidth]{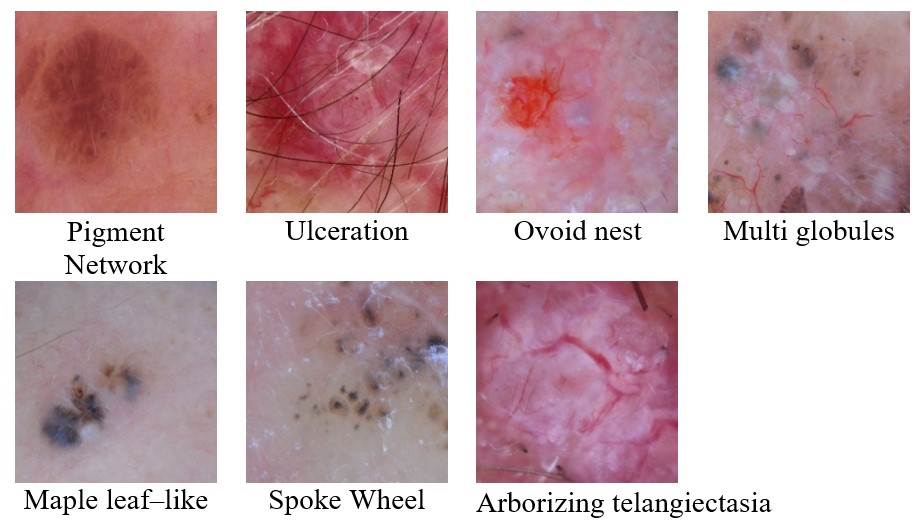}
            \caption{Examples of BCC dermoscopic criteria, whereas Pigment Network is a negative criterion}
            \label{fig:1}
        \end{figure}

        Two databases have been used in this study. The first database contains 204 dermoscopic images (256×256 pixels) from the Hospital Universitario Virgen Macarena of Seville, Spain. Each image can present one or more BCC patterns, that is, we face a multilabel classification problem. Four dermatologists label each image independently (Figure \ref{fig:1}). The number of images with a given dermoscopic BCC pattern according to Dermatologist 1 is: 16 pigment network (PN) (negative criterion), 72 ulceration (UL), 63 blue-gray ovoid nests (BO), 33 multiple blue-gray globules (MG), 28 maple leaf (ML), 12 spoke-wheel (SW), 102 arborizing telangiectasia (AT). This database will be used for the concordance analysis among different ground truths.
        The second database contains a total of 1230 dermoscopic images (256×256 pixels) from the Hospital Universitario Virgen Macarena of Seville, Spain, and its clinical area of influence (primary health centers). This database was used as a training tool for the AI tool. In this database, 588 images contain one or more BCC pattern, 577 images contain only pigment network patterns and 65 images do not present any pattern. This database was also labelled by the four dermatologists. The first database (204 images) was used as test set for the tool. 
        This study was performed in line with the principles of the Declaration of Helsinki. Approval was granted by the Ethical Committee for Biomedical Research from Andalusia (protocol code: 1901-N-22).
        
        \subsection{Interrater agreement}
        Labels are coded as binary words, where a 1/0 represents presence/absence of a certain dermoscopic pattern. The word length is seven, that is, the number of different criteria and the order is [PN, UL, BO, MG, ML, SW, AT].
        
        To assess inter-rater agreement, three different parameters are calculated:
        \begin{enumerate}
            \item Hamming distance \cite{wegner1960technique} is computed for each pair of raters to measure the discrepancy between them. This parameter is used in coding theory to calculate the number of positions where the corresponding symbols are different. For each image and dermatologist, we obtain a code word. Each code word has seven bits, one for each BCC dermoscopic criterion. A given bit is 1 if a given dermatologist has considered that the image contains this BCC pattern. Thus, if there is agreement between two dermatologists, the two code words corresponding to a given image are equal and these two dermatologists will have a Hamming distance of zero. The advantage of adapting this measure to this problem is that it allows to analyze the concordance between dermatologists in the detection of all BCC features as a whole, whereas Fleiss and Cohen-Kappa coefficients can only analyze the concordance for each feature.
        
            \item The Cohen’s Kappa coefficient \cite{CohenKappaCoefficient} is a coefficient that indicates the concordance between two raters. A Cohen’s Kappa value close to 1 indicates excellent concordance; a value close to 0 indicates random concordance; and a negative value indicates a huge discrepancy. 
        
            \item  Fleiss’ Kappa coefficient \cite{fleiss1973equivalence} is a further development of Cohen’s Kappa. It is calculated to analyze three or more raters. In our case, a Fleiss’ Kappa coefficient is obtained for each dermoscopic criterion.
        \end{enumerate}
        
        \subsection{Ground truth inference}
        A properly labeled database is crucial for the success of any machine learning system. When there is little consensus among raters, it can be difficult to determine which label should be considered GT. Different methods have been proposed to infer GT from multiple raters in the literature \cite{zhang2016learningCrowdsourced, mccluskey2021findingGT}.
        
        The most common technique for inferring the GT from multiple raters is majority voting, where the final prediction is the majority of the raters’ vote. This method assumes that each labeler has the same label quality. Since in this study the number of raters is even, in case of a tie, the dermatologist with more years of experience has the casting vote.
        
        Expectation maximisation (EM) is a powerful probabilistic method for estimating true values from noisy annotations or, in order words, for inferring the GT from multiple labelers. Dawin and Skene \cite{dawid1979ExpectationMaximization} proposed an algorithm based on expectation maximization that interactively infers the GT in two steps:
        \begin{enumerate}
            \item E-step: the probability of a particular image to have a given pattern is estimated. 
            \item M-step: the confusion matrix of each labeler and the prior probability of each class are updated.
        \end{enumerate}
        
        In this paper, both methods are compared.
        
        \subsection{Artificial Intelligence tool performance with different GTs}
        In Serrano et al., the authors presented a work that detects the BCC dermoscopic patterns present in a lesion. It is based on deep neural networks and Color Science findings \cite{serrano2022}.
        
        The same architecture has been used in this work. The training set consisted of 1230 images, which after applying data augmentation was increased to 4920 images. 
        To analyze how GT affects the performance of the AI tool, this system was trained with two different GTs independently. First, the AI tool was trained with the GT provided by Dermatologist 1 (AI tool 1). Second, the GT was inferred with the expectation maximization method as a consensus of the four dermatologists and used for a different training of the AI tool (AI tool 2). 
        
        To assess whether there are significant differences between the behavior of the two AI tools, two analyses were performed. First, McNemar’s \cite{mcnemar1947note, hoffman2019categorical} test was applied to the output of the two AI tools when they classified the database with 204 images. McNemar’s test evaluates the statistical difference between two paired binary distributions by calculating the number of disagreements between them. The following formula is evaluated:
        \begin{equation}
            \chi ^2 = \frac{(|b-c| -1)^2}{b+c},
        \label{ec:1}
        \end{equation}
        where b represents the number of disagreements when one AI tool classifies an image as positive, and c represents the number of disagreements when the other AI tool classifies an image as positive. McNemar noted that equation (1) follows a $\chi ^2$ distribution and, therefore, its value can be employed to evaluate the null hypothesis.
        
        Second, the Hamming distance between the outputs of the two AI tools was calculated when they classify the database of 204 images.
        
        \section{Results}
            \subsection{Concordance among physicians}
            Concordance among physicians is evaluated from three different points of views. First, the diagnostic concordance in BCC diagnosis is analyzed. Secondly, the agreement of their diagnosis with the biopsy is measured. Finally, the concordance in the determination of the BCC patterns present in the lesion is evaluated.
                \subsubsection{Concordance in the diagnosis of lesions}
                To analyze the concordance among dermatologists in the diagnosis of the lesions as BCC or non-BCC, two different parameters were calculated. First, Cohen-Kappa was calculated between every pair of dermatologists. The average Cohen-Kappa value was 0.9075, with a standard deviation equal to 0.0555. Cohen-Kappa concordance level according to Landis et al. \cite{landis1977measurement} is: kappa >0.8 means\textit{ almost perfect agreement}; >0.6 means \textit{substantial agreement}; >0.4 means \textit{moderate agreement}; >0.2 means \textit{fair agreement}; >0 means \textit{slight agreement}; <0 means \textit{no agreement}. Thus, the average value obtained indicates almost perfect agreement, and, as can be observed in the standard deviation, \textit{almost perfect agreement} was also obtained for every pair of raters.
                
                Secondly, Fleiss Kappa was calculated to measure the concordance among all dermatologists, which was 0.9079. According to Landis et al \cite{landis1977measurement} these values of Kappa statistics can be described as \textit{almost perfect agreement}.
        
                \subsubsection{Concordance of dermatologists' diagnosis with biopsy}
                Of the 204 lesions included in this study, 91 were biopsied. All of them were diagnosed as BCC by all dermatologists. The comparison of clinical judgment with biopsy can be summarized as follows: 88 lesions were BCC according to biopsy; 3 lesions were not BCC according to biopsy, i.e., they were false positives. Thus, the Positive Predictive Value attained by the four dermatologists was 0.967.
        
                \subsubsection{Concordance in the determination of BCC dermoscopic patterns}
                To calculate the concordance among dermatologists, Hamming distances were calculated between each pair of dermatologists (between the binary words representing their assessments). The Hamming distance has the advantage of providing a global view of the agreement between two specialists rather than analyzing the agreement only for a specific pattern. A Hamming distance equal to zero means that the binary words are equal, that is, there is full agreement between the two dermatologists. When a Hamming distance has a value of X, it means that the binary words differ in X locations. Figure \ref{fig:2} (a) shows the Hamming distances obtained for each pair of dermatologists. The information contained in this figure is summarized in Table \ref{Table:1}. We can see that the concordance is adequate when comparing Dermatologists 1, 2 and 3 (specially 1 and 3). On average, they differ in less than one BCC dermoscopic pattern; and 75.5\% of the lesions differ in at most one pattern. However, there is an important discrepancy when Dermatologist 4 is involved. This reveals the importance of inferring an appropriate GT from different physicians rather than training an AI tool from the GT given by a single physician.
               
                Table \ref{Table:2} shows the Cohen-Kappa coefficient values for each pair of dermatologists. Each column shows the agreement between each pair of raters for a dermoscopic criterion. According to Fleiss et al. \cite{fleiss2013statistical}, for most purposes, values greater than 0.75 can be taken to represent excellent agreement beyond chance, values below 0.40 may be taken to represent poor agreement beyond chance, and values between 0.40 and 0.75 may be taken to represent fair to good agreement beyond chance. Therefore, values below 0.4 are highlighted in bold. 
                As can be observed, there is a low agreement between Dermatologist 4 and the other dermatologists. 
                Table \ref{Table:2} also shows Fleiss Kappa coefficient obtained for each dermoscopic criterion. These results show a low concordance in the assessment of spoke-wheel. The overall assessment in the pigment network is also low. However, if we analyze the agreement between pair of raters, one can conclude that the low overall agreement is mainly due to rater 4. 
            \subsection{Ground Truth inference}
            This subsection aims to compare the GT obtained from a single dermatologist with the GTs obtained from a consensus of dermatologists. This is essential, as the absolute GT is not available to diagnose dermoscopic patterns in lesions. 
        
\begin{figure}[ht]
            \centering
            \includegraphics[width = \textwidth]{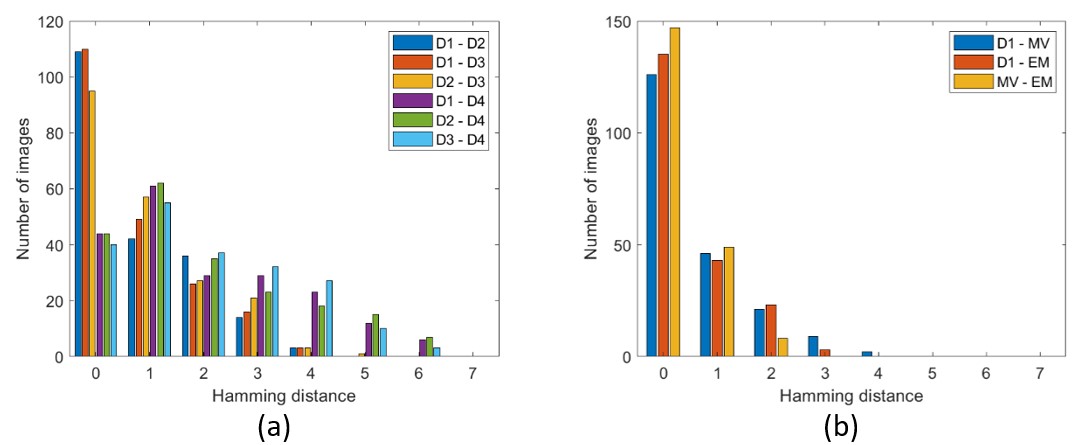}
            \caption{Comparison of the Hamming distances: (a) for every pair of dermatologists. (b) between the different GTs}
            \label{fig:2}
        \end{figure}

            Hamming distances between the different GTs are calculated and shown in Figure \ref{fig:2} (b). As can be observed, the smallest discrepancies occur between the majority voting and expectation maximization GTs, demonstrating that both methods result in similar GT inferences. 
            In comparison with Figure \ref{fig:2} (a), its results evident that inferred GTs contribute to reducing the discrepancy between the pair of raters in determining the BCC patterns present in a lesion, as both the average and the standard deviation of the Hamming distance are evidently lower between the pair of inferred GTs than between the pair of dermatologists. Table \ref{Table:1} summarizes these results.
        
            \subsection{Performance Artificial Intelligence tool with different GTs}
            To show the influence of GT on the performance of the AI tool, the AI tool was trained with two different GTs. The first GT was provided by a dermatologist (D1). The second GT was determined collectively from all dermatologists; specifically, the inferred GT of the EM was used. The same neural network architecture8 with the same hyperparameters was trained with these two different GTs. 
        
            To analyze the behavior of the AI tool trained with the two different GTs, McNemar’s test was first applied. Table \ref{Table:3} summarizes the results. In this table, it is worth noting the high number of discrepancies between the two AI tools. In particular, in the case of blue-gray ovoid nests and arborizing telangiectasia patterns, the ratio between the number of agreements and disagreements is 0.23 and 0.15, respectively. This means that, in the case of blue-gray ovoid pattern, for every 4 agreements, the two AI tools disagree on 1 image. This ratio is also high for ulceration, with a value of 0.13.
            
            As can be seen, the maximum discrepancies occur for blue-gray ovoid nests and arborizing telangiectasia patterns. Thus, according to McNemar's test, the null hypothesis between the results of both AI tools cannot be accepted for these two patterns. 
            
            Table 3 also summarizes the analysis of the Hamming distances between the output of the two AI tools, each trained with a different GT. As shown in this table, both AI tools detect the same BCC patterns in only 55.5\% of the test images. In almost 13\% of the test images, both AI tools differ in detecting two or more BCC patterns. 
            The AI tool achieves an accuracy of 96.7\% in classifying lesions into BCC or non-BCC but this accuracy decreases to 82\% when detecting dermoscopic patterns.
        
        \section{Discussion}
        In recent years, teledermatology has been widely implemented. Although very useful, this tool represents a significant workload for the Dermatology Services of public hospitals.
        
        For this reason, it would be very useful for health services to have a triage tool, which would establish a prioritization of the different cases sent from primary care, so that those cases with a higher degree of malignancy assigned could be evaluated more urgently by specialists.
        Many AI tools have been proposed in the literature to classify skin lesions, but a limited number have tried to explain the classification. Serrano et al. \cite{serrano2022} present an attempt to explain the classification of skin lesions into BCC or non-BCC, where the different BCC dermoscopic criteria present in an image are determined. However, GT on the presence/absence of these dermoscopic criteria cannot be established by biopsy. Therefore, this GT should be established from multiple dermatologists, and an in-depth study should be undertaken to infer the GT from them.
        
        In this paper, the diagnosis of four dermatologists has been collected and the GT has been inferred using two methods. Majority Voting with a casting vote to resolve ties and a method based on Expectation Maximization.
        This study leads to the conclusion that there is high concordance among dermatologists in determining whether a lesion is BCC or not, and this diagnosis is consistent with biopsy. However, there is a poor agreement among dermatologists on various dermoscopic criteria. For example, there is a low concordance in the determination of the spoke-wheel pattern. One explanation for this fact could be that only 14 out of 204 lesions present this pattern. Therefore, the probability of agreement by chance in 0 answer is very high, which tends to diminish the value of Kappa parameter. Another explanation is that the spoke-wheel pattern often appears along with other BCC patterns. Dermatologists tend to find clear signs of BCC and they may miss other subtle signs, which are not essential for diagnosis.
        
        Therefore, care must be taken when training an AI tool to determine the BCC patterns present in a lesion. Thus, a proper GT inference should be performed before training and testing AI tools. 
        The results evidence that the inferred GTs obtained with expectation maximization and with majority voting achieve a lower variability between them than between each pair of dermatologists. Specifically, the inferred GTs agree in 200 out of 204 images, while each pair of dermatologists agrees on 128 out of 204 images on average. This indicates that the consensus smoothes the outliers.
        
        Other results highlight that the AI tool is sensitive to the GT used for training. Specifically, when the AI tool was trained with D1 and with the inferred GT, in almost 13\% of the test images, both AI tools differed in detecting two or more BCC patterns.

        \section{Limitations}
        This work has several limitations. First, the database is strongly unbalanced. From 204 images, only 12 present the spoke-wheel pattern, while 102 present the arborizing telangiectasia pattern. Second, although the AI tool attains an accuracy of 96.7\% in the classification of lesions into BCC or non-BCC, it obtains an 82\% of accuracy when detecting dermoscopic patterns. This accuracy should be improved to provide a better explanation of the diagnosis.

        \section{Conclusions}
        The aim of this work is to contribute to the development of a tool to aid in the diagnosis of BCC. For this tool to be really useful, it is necessary to explain the classification it offers. However, it is difficult to develop the ability to explain its classification without establishing an adequate consensus among dermatologists' criteria.
        
        \section{Acknoledgments}
        This work was supported by the Andalusian Regional Government (PROYEXCEL\_00889),
        MCIN/AEI/10.13039/501100011033 (Grant PID2021-127871OB-I00), and ERDF/European Union (NextGenerationEU/PRTR).
        
        \clearpage
        
        \bibliographystyle{unsrt}
        \bibliography{references}  
        
        \clearpage
        
        \section{Tables and Figures}
        
        \begin{table}[ht]
            
            \caption{Summary of the information provided by the Hamming distance. Column 2: Average Hamming distance attained between every two dermatologists. Column 3: Standard deviation of the Hamming distance. Column 4: Proportion of images with Hamming distance lower than 1}
            \label{Table:1}
            \centering
            \begin{tabular}{cccc}
                \toprule
                \textbf{Pair of raters} & \textbf{Mean distance} & \textbf{STD} & \textbf{Proportion of almost agreement} \\
                \midrule
                D1-D2 & 0.8235 & 1.0424 & 151/204 (74.02\%) \\
                D1-D3 & 0.7892 & 1.0336 & 159/204 (77.94\%) \\
                D2-D3 & 0.9363 & 1.1030 & 152/204 (74.51\%) \\
                D1-D4 & 1.9314 & 1.6584 & 105/204 (51.47\%) \\
                D2-D4 & 1.9118 & 2.0237 & 106/204 (51.96\%) \\
                D3-D4 & 1.9657 & 1.5542 & 95/204 (46.57\%) \\
                \midrule
                D1-MV & 0.5588 & 0.8895 & 177/204 (86.76\%) \\
                D1-EM & 0.4363 & 0.7208 & 178/204 (87.25\%) \\
                MV-EM & 0.2500 & 0.4895 & 200/204 (98.03\%) \\
                \bottomrule   
            \end{tabular}
        
        \end{table}

        \begin{table}[ht]
            \centering
                \caption{Cohen-Kappa value calculated for every pair of physicians and Fleiss’ Kappa value estimated from all the physicians for the seven dermoscopic criteria. PN: pigment network (negative criteria), UL: ulceration, BO: blue-gray ovoid nests, MG: multiple blue-gray globules, ML: maple leaf, SW: spoke-wheel, AT: arborizing telangiectasia.}
                \label{Table:2}
                \begin{tabular}{cccccccc}
                    \toprule
                    \textbf{Cohen-Kappa} \\
                    \midrule
                    \textbf{Pair of raters} & PN & UL & BO & MG & ML & SW & AT \\
                    \midrule
                    D1-D2 & 0.6609 & 0.7100 & 0.5114 & 0.5278 & 0.4902 & 0.2031 & 0.7941 \\
                    D1-D3 & 0.7544 & 0.6989 & 0.6319 & 0.6023 & 0.4085 & 0.2110 & 0.7843 \\
                    D2-D3 & 0.5907 & 0.6353 & 0.4464 & 0.5278 & 0.3457 & 0.4262 & 0.7949 \\
                    D1-D4 & 0.1079 & 0.4808 & 0.5764 & 0.2537 & 0.2542 & 0.1958 & 0.5490 \\
                    D2-D4 & 0.1079 & 0.4764 & 0.4955 & 0.3080 & 0.2685 & 0.2332 & 0.5534 \\
                    D3-D4 & 0.0720 & 0.3835 & 0.6875 & 0.2300 & 0.2108 & 0.3475 & 0.5537 \\
                    \midrule
                    \multicolumn{1}{c}{\textbf{Fleiss’ Kappa}} \\
                    \midrule
                     & 0.1416 & 0.5625 & 0.5568 & 0.3731 & 0.2847 & 0.2559 & 0.6709 \\
                    \bottomrule
                \end{tabular}
        
        \end{table}
        
        \begin{table}[ht]
            \centering
                \caption{Discrepancy in the performance of the AI tool when trained with different GTs, according to McNemar test and Hamming distance. PN: pigment network (negative criteria), UL: ulceration, BO: blue-gray ovoid nests, MG: multiple blue-gray globules, ML: maple leaf, SW: spoke-wheel, AT: arborizing telangiectasia.}
                \label{Table:3}
                \begin{tabular}{lcccccccc}
                    \toprule
                     & PN & UL & BO & MG & ML & SW & AT \\
                    \midrule
                    Number of agreements & 166 & 160 & 146 & 176 & 171 & 176 & 156 \\
                    Number of disagreements & 14 & 20 & 34 & 4 & 9 & 4 & 24 \\
                    Patterns only detected by AI trained by D1 & 6 & 7 & 6 & 4 & 3 & 1 & 7 \\
                    Patterns only detected by AI trained by EM & 8 & 13 & 28 & 0 & 6 & 3 & 17 \\
                    \(\chi^2\) value according to McNemar test & 0.0714 & 1.2500 & 12.9706 & 2.2500 & 0.4444 & 0.2500 & 3.3750 \\
                    \(\chi^2\) probability & 0.7893 & 0.2636 & 0.0003 & 0.1336 & 0.5050 & 0.6171 & 0.0662 \\
                    \midrule
                    \textbf{Hamming distance} & 0 & 1 & 2 & 3 & 4 & 5 & 6 & 7 \\
                    \midrule
                    \% of images & 55.5\% & 31.7\% & 10.0\% & 2.2\% & 0.6\% & 0\% & 0\% & 0\% \\
                    \bottomrule
                \end{tabular}
        
        \end{table}    

\end{document}